\documentclass[journal]{IEEEtran}

\usepackage{ifpdf}
\usepackage{cite}
\usepackage{amsmath}
\usepackage{algorithm,algorithmicx}
\usepackage{threeparttable}
\usepackage[noend]{algpseudocode}
\usepackage{algorithmicx}
\usepackage{array}
\usepackage{epsfig}
\usepackage{graphicx}
\usepackage{subfigure}
\usepackage{epstopdf}
\usepackage{amssymb}
\usepackage{array}
\usepackage{xcolor}
\usepackage{booktabs}
\usepackage{multirow}
\usepackage{verbatim}

\begin{document}

\title{DV-Hop localization based on Distance Estimation using Multinode and Hop Loss in WSNs}

%Multi-Objective Localization Model based on Multinode Joint Distance Estimation 

\author{Penghong~Wang,  Xingtao~Wang, Wenrui Li,
	Xiaopeng~Fan,~\IEEEmembership{Senior Member,~IEEE,} and Debin~Zhao,~\IEEEmembership{Member,~IEEE}
        
        % <-this % stops a space
        
\thanks{This work was supported in part by the National Key Research and Development Program of China (2021YFF0900500), and the National Science Foundation of China (NSFC) under grant 61972115.}

\thanks{Penghong Wang, Xingtao Wang and Wenrui Li are with the School of Computer Science, Harbin Institute of Technology, Harbin 150001, China. e-mail: (phwang@hit.edu.cn; xtwang@hit.edu.cn; 21B903007@stu.hit.edu.cn). }

\thanks{Xiaopeng Fan and Debin Zhao are with the School of Computer Science, Harbin Institute of Technology, Harbin 150001, China, and also with the PengCheng Lab, Shenzhen 518055, China. e-mail: (fxp@hit.edu.cn; dbzhao@hit.edu.cn). }% <-this % stops a space

\thanks{Corresponding author: Xiaopeng Fan.} 
}

\markboth{}%IEEE Transactions on Vehicular Technology
{Shell \MakeLowercase{\textit{et al.}}: Bare Demo of IEEEtran.cls for IEEE Journals}

\maketitle

\begin{abstract}

Location awareness is a critical issue in wireless sensor network applications. For more accurate location estimation, the two issues should be considered extensively: 1) how to sufficiently utilize the connection information between multiple nodes and 2) how to select a suitable solution from multiple solutions obtained by the Euclidean distance loss. In this paper, a DV-Hop localization based on the distance estimation using multinode (DEMN) and the hop loss in WSNs is proposed to address the two issues. In DEMN, when multiple anchor nodes can detect an unknown node, the distance expectation between the unknown node and an anchor node is calculated using the cross domain information and is considered as the expected distance between them, which narrows the search space. When minimizing the traditional Euclidean distance loss, multiple solutions may exist. To select a suitable solution, the hop loss is proposed, which minimizes the difference between the real and its predicted hops. Finally, the Euclidean distance loss calculated by the DEMN and the hop loss are embedded into the multi-objective optimization algorithm. The experimental results show that the proposed method gains 86.11\% location accuracy in the randomly distributed network, which is 6.05\% better than the DEM-DV-Hop, while DEMN and the hop loss can contribute 2.46\% and 3.41\%, respectively.

\end{abstract}


\IEEEpeerreviewmaketitle

\section{Introduction}

\IEEEPARstart{W}{ireless} sensor networks (WSNs) have attracted much attention in many fields\cite{lv2019infrastructure02, gungor2010opportunities03, 5261926Lakafosis}, e. g. industry, agriculture, medical treatment, and military. In such applications, the localization of sensor nodes is a crucial research issue. Especially, WSNs are composed of a group of tiny sensor nodes equipped with small batteries, which have the characteristics of low energy consumption and low cost. This low-cost requirement limits many range-based sensor node location methods \cite{Klukas002, niculescu2003ad04, girod2002locating05, tiemann2017scalable06}, as the range-based localization method relies on the physical characteristics (such as signal strength or arrival time) of the received signal to measure the distance between sensor nodes. This localization methods increase energy consumption and localization cost of sensors. On the contrary, the range-free localization methods \cite{niculescu2003dv09, vivekanandan2007concentric07, 5358126Liu, 5601108Bao, 4370265Nan, chuan2008research015, Xiao2010,  Bang2012A,  Slim2016, guadane2017enhanced016, chen2020smart018, yu2006improved011, shahzad2016dv012, shi2018distance013, gui2020connectivity014, cui2017novel023, singh2018pso019, cui2018high020, cai2019research021, huang2019weighted017, liu2019improved022, cai2020weight010, 8913604@023_1, OUYANG@023_2, kanwar2020multiobjective026, cai2019multi024, wang2020gaussian025, kanwar2021range027, chen2022enhanced, jin2022novel, Qiang2023, Cao2023, TVTWang77} can obtain the locations of unknown nodes without additional hardware.

\begin{figure}[t]
	
	\center{\includegraphics[scale=0.7]{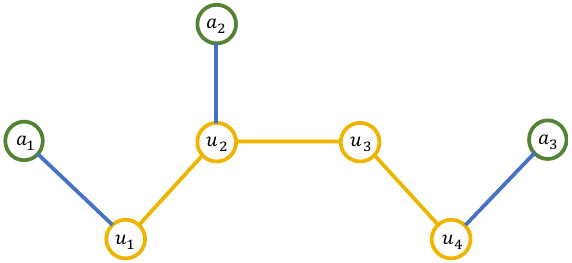}}
	\caption{An illustration of WSNs. Anchor nodes ($ a_1, a_2 $ and $ a_3 $) and unknown nodes ($ u_1, u_2, u_3,  $ and $ u_4 $).}
	\label{BG}
\end{figure}

Range-free localization schemes generally predict the location of unknown nodes based on the connectivity among sensor nodes. Fig. \ref{BG} is an illustration of WSN, which contains seven sensor nodes, of which three are anchor nodes (i.e., nodes with a known location, recorded as $ a_1 $, $ a_2 $ and $ a_3 $) and four are unknown nodes (i.e., nodes with an undisclosed location, recorded as $ u_1, u_2, u_3 $ and $ u_4 $). The edge between two nodes indicates that these two nodes are neighbors, i.e., they can communicate with each other and vice versa. Based on the locations of $ a_1 $, $ a_2 $ and $ a_3 $ and the hops to each unknown node ($ u_1, u_2, u_3 $ or $ u_4 $), the locations of these unknown nodes can be predicted.

Distance vector hop (DV-Hop) is a typical range-free localization algorithm proposed by Niculescu and Nath \cite{niculescu2003dv09}. Its positioning process includes three stages. \textbf{1)} Minimum hops statistics: record the minimum connection hops from an anchor node to other nodes; \textbf{2)} Distance estimation: calculate the distances per hop of anchor nodes based on these geographical locations and the minimum connection hops between them, and calculate the distance between an unknown node and an anchor node based on the minimum connection hops and distance per hop of anchor node; \textbf{3)} Location prediction: predict the geographical location of each unknown node based on the obtained information.

Most of the existing range-free localization works draw on the experience of the DV-Hop scheme, such as \cite{ 5358126Liu, 5601108Bao, 4370265Nan, chuan2008research015, Bang2012A, guadane2017enhanced016, chen2020smart018, yu2006improved011, shahzad2016dv012, Slim2016, shi2018distance013, gui2020connectivity014, cui2017novel023, singh2018pso019, cui2018high020, cai2019research021, huang2019weighted017, liu2019improved022, cai2020weight010, 8913604@023_1, OUYANG@023_2, kanwar2020multiobjective026, cai2019multi024, wang2020gaussian025, kanwar2021range027, Liu2020, Xiao2010,   chen2022enhanced, jin2022novel, TVTWang77} proposing various improvement strategies in the stage of distance calculation and position prediction. Specifically, \cite{4370265Nan, cui2017novel023, singh2018pso019, cui2018high020, cai2019research021, huang2019weighted017, OUYANG@023_2} use a single-objective intelligent optimization algorithm for location prediction.  \cite{shahzad2016dv012, cai2020weight010, kanwar2020multiobjective026, cai2019multi024, wang2020gaussian025, kanwar2021range027,TVTWang77} develop a multi-objective optimization model to predict the locations of unknown nodes. \cite{Bang2012A, Slim2016, liu2019improved022, TVTWang77, cai2020weight010, wang2024probabilitybased} develop distance estimation strategies between an unknown node and an anchor node.

In these works, the following two issues should be considered extensively: 1) how to sufficiently utilize the connection information between multiple nodes and 2) how to select a suitable solution from multiple solutions obtained by Euclidean distance loss. To address the two issues, a DV-Hop localization based on distance estimation using multinode (DEMN) and the hop loss in the WSNs is proposed. Specifically, the DEMN strategy aims to leverage the information constraints provided by multiple anchor nodes to narrow the potential search space of an unknown node and improve the accuracy of estimated distance. The hop loss can help the algorithm select a suitable solution from the set of potential solutions obtained in iterative calculations. These schemes offer a promising solution to enhance the localization accuracy in WSNs. The main contributions can be summarized as follows:

\begin{enumerate}
	\item [\textbf{1)}] We propose DEMN, which predicts the distance from an unknown node to an anchor node using multinode information. Specifically, when multiple anchor nodes can detect an unknown node, the distance expectation between the unknown node and an anchor node is calculated using the cross domain information and is considered as the expected distance between them, which narrows the search space.

	\item [\textbf{2)}] We propose the hop loss, which minimizes the difference between the real and its predicted hops. Hop loss enables us to choose the most suitable solution as the output solution when there are multiple potential solutions.

	\item [\textbf{3)}] We embed the Euclidean distance loss and the hop loss into a multi-objective genetic algorithm. The simulation results show that the proposed method gains 86.11\% location acuracary in the randomly distributed network, which is 6.05\% better than the DEM-DV-Hop, while the DEMN and the hop loss can contribute 2.46\% and 3.41\%, respectively.

\end{enumerate}

The rest of this paper is organized as follows: Section II
reviews the related works on the DV-Hop algorithm. Section III details the distance estimation using multinode (DEMN). Section IV presents the hop loss. Section V introduces the multi-objective optimization embedding the Euclidean distance loss calculated by the DEMN and the hop loss. Section VI provides the simulation results. Finally, the conclusion of this paper is summarized.

\section{Related Work}

Most range-free localization algorithms need to obtain the necessary information to predict the geographical locations of unknown nodes, such as the locations of anchor nodes and the connectivity between nodes. The DV-Hop algorithm is proposed by Niculescu \textit{et al.} \cite{niculescu2003dv09} provides a simple calculation scheme, and its process is as follows. Firstly, beacon nodes disseminate data packets and flood the entire network. Thereinto, the structure of the package is \{$ ID, L_{i},Hop_{i} $\}, including the ID, locations of anchor nodes, and initial hop ($ Hop_{i}=0 $) information. When other nodes receive the packet, they will update the $ Hop_{i} $ in the package and save the copy to their table. When the flooding process converges, each node has a table containing the minimum hop count to other nodes. Then, based on the geographical location of anchor nodes and the minimum number of hops between nodes, the average distance ($ Avg\_DisHo{p_i} $) per hop of an anchor node is calculated by
\begin{equation}
Avg\_DisHo{p_i} = \frac{{\sum\limits_j {\sqrt {{{({x_i} - {x_j})}^2} + {{({y_i} - {y_j})}^2}} } }}{{\sum\limits_j {Ho{p_{i,j}}} }},
\label{eq0}
\end{equation}
where ${({x_i}, {y_i})}$ and ${({x_j}, {y_j})}$ represent the locations of node $ {a}_i $ and node $ {a}_j $ respectively, $ {Hop}_{i, j} $ denotes the minimum hop count between  $ {a}_j $ and $ {a}_i $. The estimated distance between $ {u}_k $ and $ {a}_i $ is calculated by
\begin{equation}
{Dis}_{i, k} = Avg\_DisHo{p_i} \cdot {Hop}_{i, k},
\label{eq01}
\end{equation} 
where $ {Hop}_{i, k}$ denotes the minimum hop count between $ {u}_k $ and $ {a}_i $. The least square method is used to predict the location of $ {u}_k $.

\begin{figure*}[htbp]
	
	\center
	\subfigure[]{\includegraphics[scale=0.7]{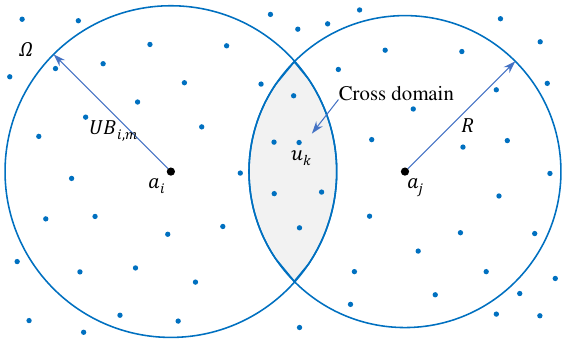}}
	\subfigure[]{\includegraphics[scale=0.7]{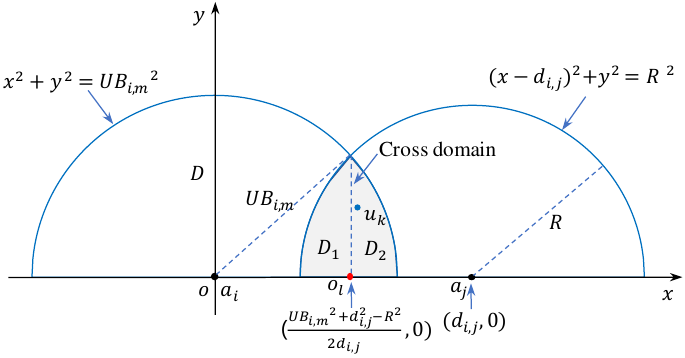}}
	
	\caption{Distance estimation using multinode. $ {a}_{i} $ and $ {a}_{j} $: two distinct anchor nodes, $ {u}_{k} $: an unknown node, $ {o}_{l} $: an ordinary node, $ D_1 $ and $ D_2 $: a part of the cross domain, $ m $: the hop count, $ R $: communication radius,  $ d_{i,j} $: the distance between node $ {a}_{i} $ and node $ {a}_{j} $.}
	\label{joint}
\end{figure*}

To improve the localization accuracy of DV-Hop, scholars have developed different improvement strategies. Specially, \cite{4370265Nan} models the location problem as a minimum optimization problem for the estimation distance and embeds the model into the intelligent optimization algorithm. \cite{cui2017novel023, singh2018pso019, cui2018high020, cai2019research021, huang2019weighted017, OUYANG@023_2} focused on improving the location prediction algorithm for unknown nodes. \cite{cai2019research021} improves the global search ability of the algorithm by introducing optimization of inertia weights and selection strategies of adaptive functions. \cite{OUYANG@023_2} proposes an adaptive strategy for crossover probability and mutation probability to improve the convergence performance of genetic algorithms. \cite{Bang2012A, Slim2016, liu2019improved022, TVTWang77, cai2020weight010, cui2018high020} improve the accuracy of the estimated distance by improving the calculation method of the distance between an unknown node and an anchor node. For example, Cui \textit{et al.}\cite{cui2018high020} believe that estimating the distance between nodes based on discrete hops in DV-Hop can result in significant errors. Therefore, when nodes follow a uniform distribution, they propose to transform discrete hop into continuous hop count. It can obtain more accurate hop count values among nodes, thereby improving the accuracy of distance estimation. Wang \textit{et al.} \cite{TVTWang77} proposed a distance estimation model based on the distribution characteristics of nodes. This model mathematically derives the farthest and average distance detected by an anchor node under different hops, improving the predicted distance's accuracy.

In recent years, scholars have also developed some new localization schemes for different localization scenarios, such as positioning sensors in irregular areas \cite{shahzad2016dv012, wang2020gaussian025, TVTWang77} or sensors with obstacles \cite{cai2019multi024, Liu2020, Xiao2010}. For example, \cite{Liu2020} judges whether there is a severe detour in the connectivity of nodes through the theoretical analysis of the minimum hops of nodes and geometric constraints. Then, the geometric constraints of anchor pairs are used to estimate the unknown node position. In addition, \cite{cai2019multi024} introduces the theoretical distance per hop of anchor nodes and propose a multi-objective localization loss models, then embed these loss models into the multi-objective optimization algorithm to predict the geographical location of each unknown node. Specially, \cite{gui2020connectivity014} takes the discrepancy
between the predicted connectivity of the nearest neighbor
node and its basic connectivity as a constraint to optimize
the expected location.

For achieving more accurate location estimation in wireless sensor networks, two critical issues need to be extensively considered: 1) how to effectively leverage the connection information between multiple nodes and 2) how to select the most suitable solution from multiple solutions obtained by the Euclidean distance loss. To address these issues, this paper proposes a novel approach called "DV-Hop localization based on the DEMN and the hop loss" for WSNs. The proposed method differs from the approach presented in \cite{gui2020connectivity014} in the following ways:
\textbf{1)} \cite{gui2020connectivity014} constructs a constraint optimization problem based on the real connectivity between any two adjacent nodes, while the hop loss considers multi-hop information between nodes.
\textbf{2)} The DEMN technique narrows the search space of unknown nodes and helps to obtain accurate estimated distances. In summary, the proposed DV-Hop localization scheme, incorporating DEMN and hop loss, addresses the key challenges of leveraging connection information and selecting appropriate solutions. These differences from \cite{gui2020connectivity014} enable the proposed method to achieve more accurate and reliable location estimation in WSNs.

\section{Distance Estimation using Multinode }

In this section, we provide a detailed introduction to the design of DEMN. The primary objective of the proposed DEMN strategy is to enhance the accuracy of the estimated distance between an unknown node and an anchor node in WSNs. It is achieved by utilizing the information constraints obtained from multiple anchor nodes, which helps to reduce the potential search space for each unknown node. As a result, the accuracy of distance estimation is significantly improved. Fig. \ref{joint}(a) illustrates the sensor nodes deployed in a 2D plane following a uniform distribution. In this figure, $a_i$ and $a_j$ represent two anchor nodes, while $u_k$ denotes an unknown node located in a cross domain that can be detected by both node $a_i$ and node $a_j$. The parameter $UB_{i,m}$ represents the upper bound detected by node $a_i$ when the hop count is $m$. The calculation of $UB_{i,m}$ is based on the approach described in \cite{TVTWang77}. Fig. \ref{joint}(b) further illustrates an example analysis of modeling these sensor nodes in a Cartesian coordinate system. This representation allows for a better understanding and visualization of the sensor node distribution and their relationships. By combining the insights from Fig. \ref{joint}(a) and (b), the DEMN strategy optimizes the distance estimation process, taking into account the constraints imposed by multiple anchor nodes and effectively narrowing down the search space for unknown nodes. The relevant information is defined as follows:

\textbf{Definition 1.} Let $ Circle(a_0, r_0) $ define a circle with the center of $ a_0 $ and the radius of $ r_0 $, its function in Cartesian coordinate system is expressed as $ \Psi_0:x^2+y^2=r_0^2 $.

\textbf{Definition 2.} Let $ a_i $ define the origin and has $ (0, 0) $ as coordinates. The coordinates of $ a_i $ are $ (d_{i,j},0) $.

Based on the above definition, $ Circle(a_i,{UB}_{i,m}) $ can be represented as $ \Psi_1:x^2+y^2={{UB}_{i,m}}^2 $, and $ Circle(a_j,R) $ can be represented as $ \Psi_2:{(x-d_{i,j})}^2+y^2={R}^2 $, where $ x $ and $ y $ are two variables. 
The coordinate $ (\frac{{UB_{i,m}}^2{+d}^2-R^2}{2d_{i,j}}, 0) $  of $ o_l $  is obtained by combining $ \Psi_1 $ and $ \Psi_2 $. The expected distance between $ u_k $ in the cross domain and $ a_i $  is considered as the expected (estimated) distance between them, and it is calculated by
\begin{equation}
\begin{aligned}
{E\_dis}_{i,k}&=\frac{\int_{d_{i,j}-R}^{\frac{{UB_{i,m}}^2{+d}^2-R^2}{2d_{i,j}}}{\int_{0}^{\sqrt{R^2-{(x-d_{i,j})}^2}}\sqrt{x^2+y^2}dydx}}{{D_1}}\\
&+\frac{\int_{\frac{{UB_{i,m}}^2{+d_{i,j}}^2-R^2}{2d_{i,j}}}^{UB_{i,m}}{\int_{0}^{\sqrt{{UB_{i,m}}^2-x^2}}\sqrt{x^2+y^2}dydx}}{{D_2}}
\end{aligned},
\label{Expdis}
\end{equation}
where $ d_{i,j} $ denotes the distance between node $ a_i $ and $ a_j $. $ {D_1} $ and $ {D_2} $ respectively refer to the area of the corresponding regin, and they are calculated by
\begin{equation}
\begin{aligned}
{D_1}=\int_{d_{i,j}-R}^{\frac{{UB_{i,m}}^2{+d}^2-R^2}{2d_{i,j}}}{\sqrt{R^2-{(x-d_{i,j})}^2}dx} \\ 
\end{aligned}, 
\label{SD1}
\end{equation}

\begin{equation}
\begin{aligned}
{{D_2}} =\int_{\frac{{UB_{i,m}}^2{+d_{i,j}}^2-R^2}{2d_{i,j}}}^{UB_{i,m}}{\sqrt{{UB_{i,m}}^2-x^2}dx} \\ 
\end{aligned}.
\label{SD2}
\end{equation}
In Eq. (\ref{Expdis}), we set $ m $ to be less than 3. When $ m\geq3 $, the expected distance follows the calculation method proposed by \cite{TVTWang77}.

\begin{algorithm}[t]
	\caption{Calculation of the $ {E\_dis}_{i,k} $.} %算法的名字
	\label{Edis}
	
	\hspace*{0.02in} {\bf Input:} $ d_{i,j} $: the distance between node $ a_i $ and node $ a_j $
	
	\hspace*{0.45in} $ hop_{i,k} $ and $ hop_{i,k} $
	
	\hspace*{0.02in} {\bf Output:} 	$ {E\_dis}_{i,k} $
	\begin{algorithmic}[1]
		% \State 后写一般语句
		
		\State Calculate $ d_{i,j} $ between node $ a_i $ and $ a_j $.
		\For{$ k=1 \to N_u $} 
		
		\For{$ i=1\to N_a, j=1\to N_a$}
		\If{$ hop_{i,k}=m $ \& $ hop_{j,k}=1$} 
		
		\State $ {{D_1}}  \gets Eq. (\ref{SD1}) $
		\State $ {{D_2}}  \gets Eq. (\ref{SD2}) $
		\State $ {E\_dis}_{i,k}  \gets Eq. (\ref{Expdis}) $

		\EndIf
		\State{\bf end}
		
		\EndFor
		\State{\bf end}
		
		\EndFor
		\State{\bf end}
		
	\end{algorithmic}
	
\end{algorithm}

The proposed DEMN strategy exhibits significant differences and advantages compared to the distance estimation model introduced in \cite{TVTWang77}. The DEMN strategy is based on multiple anchor nodes to narrow the potential search space of an unknown node and enhance the accuracy of estimated distance. As depicted in Fig. \ref{joint}(b), assuming $ m=1 $, the previous work \cite{TVTWang77} defines the potential search space of an unknown node as the semicircular region $ D $ covered by $ Circle(a_i, UB_{i,1}) $. In contrast, the proposed DEMN defines it as the region $ D_1+D_2 $. This is a crucial improvement that significantly narrows the potential search space for an unknown node.

Algorithm \ref{Edis} provides the implementation details of the $ {E\_dis}_{i,k} $. After obtaining the $ {E\_dis}_{i,k} $, we used it to construct a Euclidean distance loss. The traditional Euclidean distance loss is expressed as
\begin{equation}
\min \;{f_1}{({x_k},{y_k})} =  \sum\limits_{i=1}^{N_{a}} {(\sqrt {{{({x_i} - {x_k})}^2} + {{({y_i} - {y_k})}^2}}  - Dis{_{i,k}})^2},
\label{f1}
\end{equation}
where ${({x_i}, {y_i})}$ and ${({x_k}, {y_k})}$ denote the locations of node $ {a}_i $ and node $ {u}_k $ respectively, $ {Dis}_{i, k} $  represents the estimated distance between node $ {u}_k $ and node $ {a}_i $. In this paper, $ {Dis}_{i, k} $ refers to $ {E\_dis}_{i,k} $ and it is calculated by Eq.(\ref{Expdis}). To reduce computational costs, the DEMN only considers the scenario where an unknown node is detected by two anchor nodes and the hop counts are $ hop_{i,k}<3$ and $ hop_{j,k}=1 $ respectively.

\section{Hop Loss}

\begin{figure}[htbp]
	
	\center
	{\includegraphics[scale=0.6]{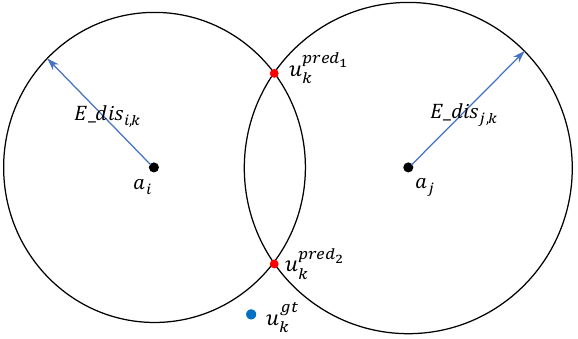}}
	
	\caption{Design principle of hop loss. $ {u}^{{pred}_{1}}_{k} $ and $ {u}^{{pred}_{2}}_{k} $: two predicted locations of node $ {u}_{k} $, $ {u}^{gt}_{l}  $: the ground truth location of node $ {u}_{l} $, $ {a}_{i} $ and $ {a}_{j} $: two distinct anchor nodes.}
	\label{feedb}
\end{figure}

Minimizing the Euclidean distance loss may obtain multiple equivalent solutions. Assuming there are two anchor nodes $ {a}_{i} $ and $ {a}_{j} $, as well as an unknown node $ {u}_{k} $, as shown in Fig. \ref{feedb}. Where $ {u}^{gt}_{k}  $ denotes the ground truth locations of node $ {u}_{k} $, $ E\_dis_{i,k}  $ and $ E\_dis_{j,k}  $ are the predicted distances from node $ {u}_{k} $ to two anchor nodes, respectively. When the Euclidean distance loss is minimal, i.e. $ f_1({x_k}, {y_k})=0 $, we obtain two predicted locations $ {u}^{{pred}_{1}}_{k} $ and $ {u}^{{pred}_{2}}_{k} $. As shown in Fig. \ref{feedb}, $ {u}^{{pred}_{2}}_{k} $ is more suitable than $ {u}^{{pred}_{1}}_{k} $. In this case, Euclidean distance loss is ineffective in selecting a suitable solution, however, randomly selecting one of the solutions may cause a serious error. To address this issue, this section proposes the hop loss function, which can choose a more suitable solution by minimizing the difference between the real hop and its predicted hop.

\begin{figure}[htbp]
	
	\centerline{\includegraphics[scale=0.83]{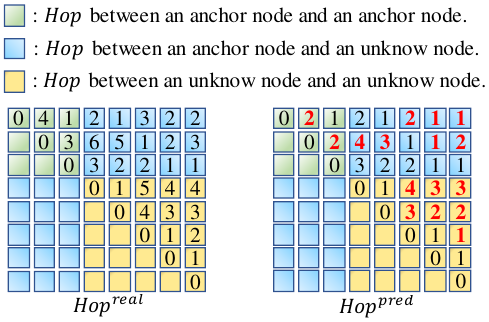}}
	\caption{An example of the real and its predicted hops. $ {Hop}^{real} $: the real hops, $ {Hop}^{pred} $: the predicted hops.}
	\label{hopm}
\end{figure}

Fig. \ref{hopm} illustrates an example of the real and its predicted hops. In the figure, the left part represents the real hop ($ {Hop}_{i,k}^{real} $) between nodes, which is obtained by the flooding broadcast of packets\cite{niculescu2003dv09}. The right part is the predicted hop ($ {Hop}_{i,k}^{pred} $) between nodes, which is calculated based on the predicted locations of nodes, as shown in Algorithm \ref{SPD}. The smaller the difference between $ {Hop}_{i,k}^{pred} $ and $ {Hop}_{i,k}^{real} $, the more closer $ {Hop}_{i,k}^{pred} $ to $ {Hop}_{i,k}^{real} $, which means the chosen solution is more accurate.
Based on the above analysis, the hop loss is proposed as
\begin{equation}
\begin{aligned}
\min \;f_2({Hop}_{i,k}^{pred})=&\sum_{k=i}^{N}{\sum_{i=1}^{N}{({Hop}_{i,k}^{real}-{Hop}_{i,k}^{pred})}^2} 
\\
&,\forall{Hop}_{i,k}^{real}<3  \\
\end{aligned}.
\label{f2}
\end{equation}

\begin{algorithm}[htbp]
	\caption{Calculation of the $ {Hop}_{i,k}^{pred} $ .} %算法的名字
	\label{SPD}
	\hspace*{0.02in} {\bf Input:} 
	Anchor node locations, $ R $ and the expected locations \hspace*{0.5in}of unknown nodes 
	
	\hspace*{0.02in} {\bf Output:} %算法的结果输出
	$ {Hop}_{i,k}^{pred} $
	\begin{algorithmic}[1]
		% \State 后写一般语句
		
		\For{$ i=1\to N, k=i \to N$}  
		
		\State Calculate the $ Euc\_Dis_{i,k}^{pred} $ between node $ o_{i} $ and $ o_{k} $ 
		
		\If{$ Euc\_Dis_{i,k}^{pred} <R$} 
		
		\State Node $ o_{i} $ and $ o_{k} $ are adjacent nodes
		\State Record their indexes [$ row,col $] 
		
		\EndIf
		\State{\bf end}
		
		\EndFor
		\State{\bf end}
		
		\State $ G \gets graph(col,row) $ 
		
		\State $ {Hop}_{i,k}^{pred} \gets BFsearch(G)$
		
	\end{algorithmic}
	
\end{algorithm}

Algorithm \ref{SPD} provides the implementation details of the $ {Hop}_{i,k}^{pred} $, where $ G $ denotes the undirected graph, $ o_{i} $ and $ o_{k} $ denote the ordinary sensor nodes, $ Euc\_Dis_{i,k}^{pred} $ denotes the predicted Euclidean distance between  node $ o_{k} $ and node $ o_{i} $. Firstly, we combine the predicted locations of unknown nodes obtained from each iteration of the algorithm with the anchor nodes locations to calculate $ Euc\_Dis_{i,k}^{pred} $. Secondly, record adjacent node pairs by comparing $ Euc\_Dis_{i,k}^{pred} $ and communication radius $ R $. Thirdly, these adjacent nodes are constructed into an undirected graph. Finally, the $ {Hop}_{i,k}^{pred} $ in the graph is obtained by the breadth-first search function, i.e., $ BFsearch(G) $.

\section{Multi-Objective Optimization}

In this section, we embed the Euclidean distance loss calculated by DEMN and the hop loss into the multi-objective genetic algorithm (NSGI-II \cite{deb2002fast031}) for iterative optimization. It is emphasized that the main contributions of this paper are as follows: 
1) \textbf{The DEMN strategy}. The proposed approach effectively utilizes the constraint information provided by multiple anchor nodes to narrow down the potential search space of an unknown node. This constraint information enhances the accuracy of predicting the distance between an unknown node and an anchor node, leading to more reliable localization.
2) \textbf{The hop loss}. The incorporation of hop loss in the optimization process enhances the ability to select suitable solutions and make the localization results more accurate and robust.
The multi-objective genetic algorithm serves as a necessary tool to solve the locations of unknown nodes. Algorithm \ref{Alg2} outlines the pipeline of a multi-objective genetic algorithm. Where more details about crossover, mutation and selection operation, non-dominated sorting, and crowding distance calculation are following \cite{deb2002fast031}.

\begin{algorithm}[htbp]
	
	\caption{Multi-objective genetic algorithm.} %算法的名字
	\label{Alg2}
	
	\hspace*{0.02in} {\bf Input:} 
	Node set: includes anchor nodes and unknown nodes 
	
	\hspace*{0.02in} {\bf Output:} %算法的结果输出
	The locations of unknown nodes
	\begin{algorithmic}[1]
		% \State 后写一般语句
		
		\State Initialize parameters with Table \ref{tab1} and initial population.
		
		\State $ {E\_dis}_{i,k}$ $\gets $ Algorithm \ref{Edis}
		
		\While{$ iter<MIter $}

		\State Perform crossover and mutation operations based on \hspace*{0.2in}$ Pc $ and $ Pm $.
		
		\State $ {Hop}_{i,\ j}^{pred}$ $\gets $ Algorithm \ref{SPD}
		
		\State $ f_1 $ $\gets $ Eq.(\ref{f1})
		
		\State $ f_2 $ $\gets $ Eq.(\ref{f2}).
		
		\State Perform non-dominated sorting and crowding distance \hspace*{0.22in}calculation based on the values of $ f_1 $ and $ f_2 $.
		
		\State Select population individuals based on sorting results.
		
		\State $ iter = iter + 1 $
		
		\EndWhile
		\State{\bf end} 
	\end{algorithmic}
\end{algorithm}

After finishing the iterations, we obtain a set of solutions \{$ Indl_1 $, ..., $ Indl_i $, ..., $ Indl_{Ps} $\}. $ Indl_i $ denotes the $ i^{th} $ individual in the population, it contains the predicted locations of all unknown nodes. Based on the predicted locations of the unknown nodes represented by $ Indl_i $, we can obtain the values of two Loss function $ f_1(Indl_i ) $ and $ f_2(Indl_i) $. Especially, we set the individual with the minimum hop loss value (i.e., min $ f_2(Indl_i) $) as the final output solution of the algorithm.

In summary, the integration of the Euclidean distance loss, hop loss, and the multi-objective genetic algorithm enables the proposed approach to iteratively optimize the localization of unknown nodes in the WSNs. The contributions of this paper lie in the effective utilization of constraint information provided by multiple anchor nodes, the introduction of hop loss, and the utilization of the multi-objective genetic algorithm to achieve more accurate and reliable localization results.

\section{Simulation And Analysis}

In this section, we provide the experimental results and the comparison with the state-of-the-art results, including DV-Hop \cite{niculescu2003dv09}, CC-DV-Hop\cite{gui2020connectivity014},  OCS-DV-Hop\cite{cui2017novel023}, CCS-DV-Hop\cite{8913604@023_1}, IAGA-DV-Hop\cite{OUYANG@023_2}, NSGAII-DV-Hop\cite{cai2019multi024} and DEM-DV-Hop\cite{TVTWang77}.

\subsection{Experimental Setup}

\begin{figure*}[t]
	
	\center
	\subfigure[]{\includegraphics[scale=0.47]{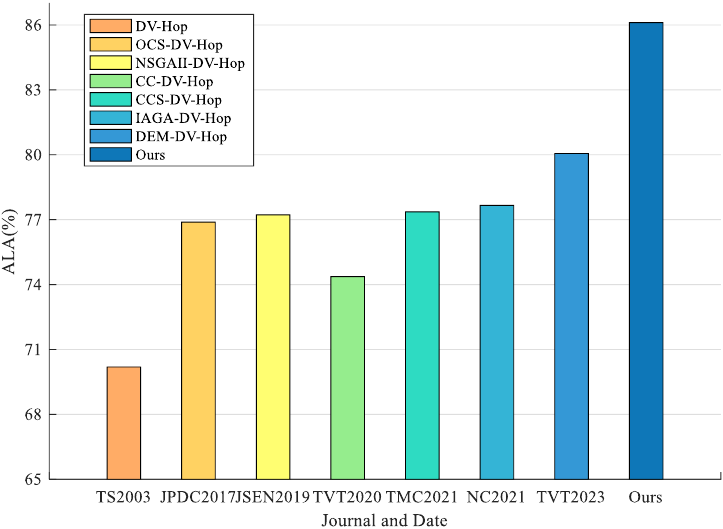}}~~~
	\subfigure[]{\includegraphics[scale=0.47]{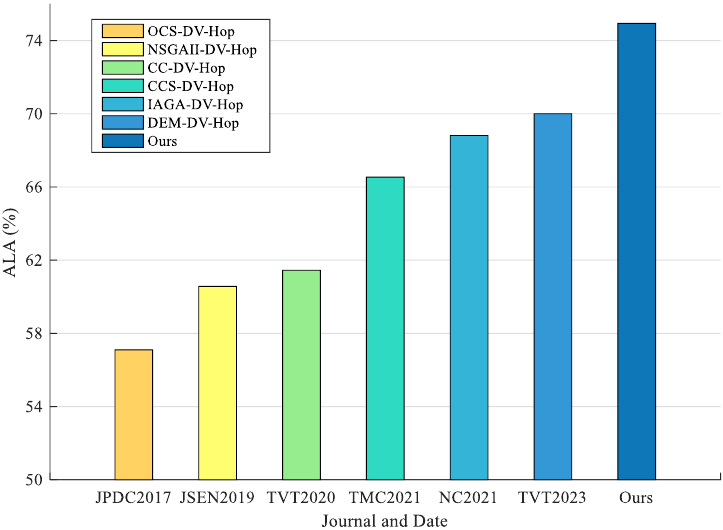}}
	\subfigure[]{\includegraphics[scale=0.47]{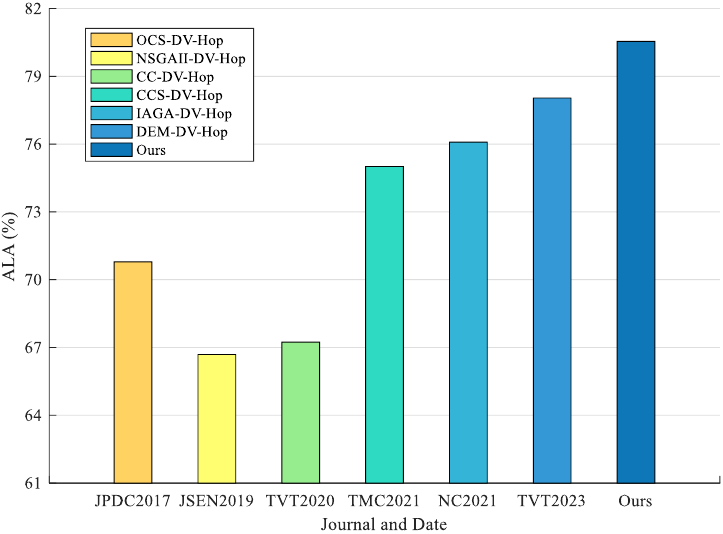}}~~~
	\subfigure[]{\includegraphics[scale=0.47]{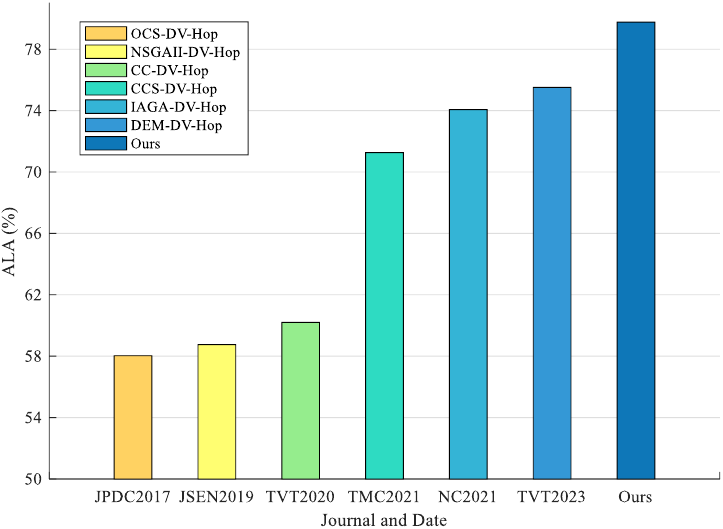}}
	
	\caption{The $ ALA $ of different algorithms. (a): randomly distributed network; (b): C-shaped network; (c): O-shaped network; (d): X-shaped network. TS: Telecommunication Systems; JPDC: Journal of Parallel and Distributed Computing; JSEN: IEEE Sensors Journal; TVT: IEEE Transactions on Vehicular Technology; TMC: IEEE Transactions on Mobile Computing; NC: Neurocomputing. }
	\label{ALA}
\end{figure*}

\textbf{Datasets.} We choose four types of node distribution data for localization performance evaluation. As shown in Fig. \ref{ALA}, the four kinds of sensor test networks are random networks and C-, O- and X-shaped networks. Each test network contains the true locations of 100 sensor nodes distributed in a 100m×100m region. These nodes' first $ 5\%-30\% $ are the anchor node locations, and the rest are unknown nodes' ground truth (gt) locations.

\begin{table}[htbp]
	\renewcommand{\arraystretch}{1.15}
	\caption{Simulation parameters. }
	\label{tab1}
	\centering
	\begin{threeparttable}
		\begin{tabular}{p{2.7cm}<{\centering}  p{1.3cm}<{\centering}   p{1.5cm}<{\centering}   p{1.3cm}<{\centering} }
			\hline
			\hline
			\textbf{ Parameters} & \textbf{Value}   &\textbf{ Parameters}		& \textbf{Value}    \\
			
			\hline
			
			$ Pc $      & 0.9 & 	$ Pm $    & 0.1   \\
			
			$ N_{u} $     & 70-95  & $ R  $ (m)   & 25-40  \\
			
			Population size       & 20  &  $ MIter $       & 500 	  \\
			
			Independent repeat test    & 50   & -- &--	\\
			
			\hline
			\hline
			
		\end{tabular}
		*$ Pc $: crossover probability,   $ Pm $: mutation probability, $ N_{u} $: the number of unknown nodes, $ MIter $: maximum iterations, independent repeat test: the number of independent experiments; population size: the number of individuals in the population.
	\end{threeparttable}
\end{table}

\textbf{Parameters.} The detailed experimental parameters of the algorithm are listed in Table \ref{tab1}. The parameter settings employed in this study are consistent with those utilized in the algorithms proposed by \textit{Cai et al.}\cite{cai2019multi024} and  Wang \textit{et al.}\cite{TVTWang77}.

\begin{figure}[t]
	\center
	\subfigure[]{\includegraphics[scale=0.35]{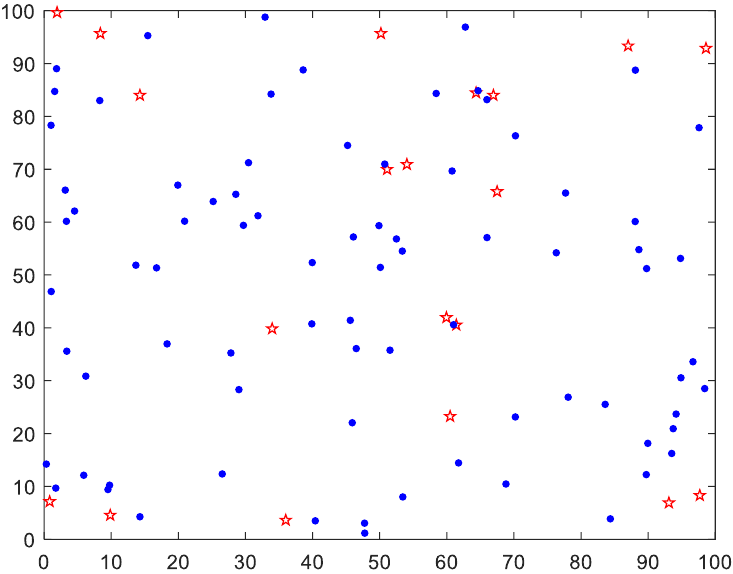}}
	\subfigure[]{\includegraphics[scale=0.35]{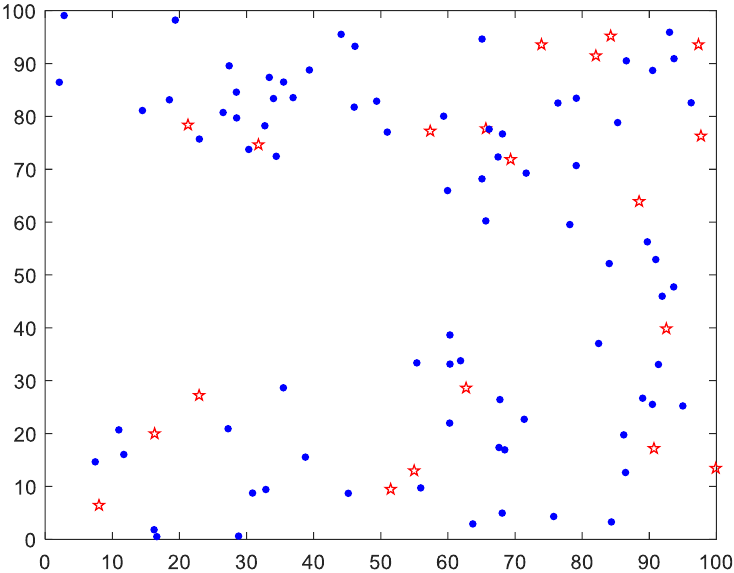}}
	\subfigure[]{\includegraphics[scale=0.35]{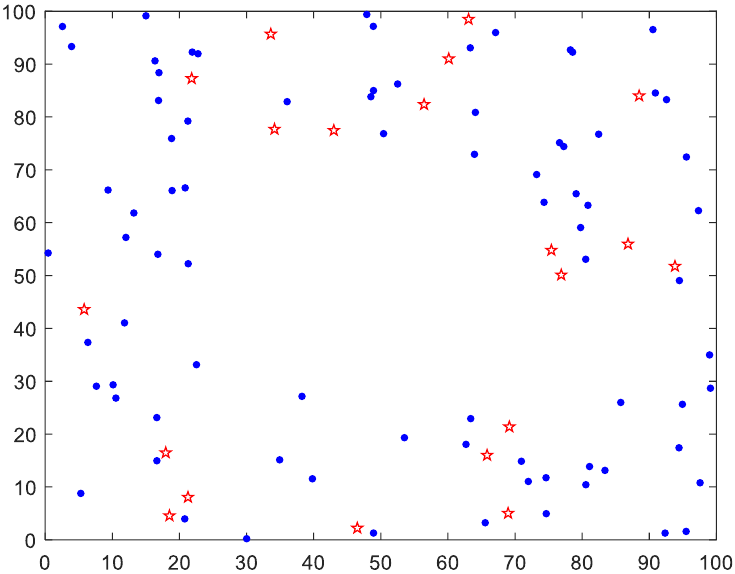}}
	\subfigure[]{\includegraphics[scale=0.35]{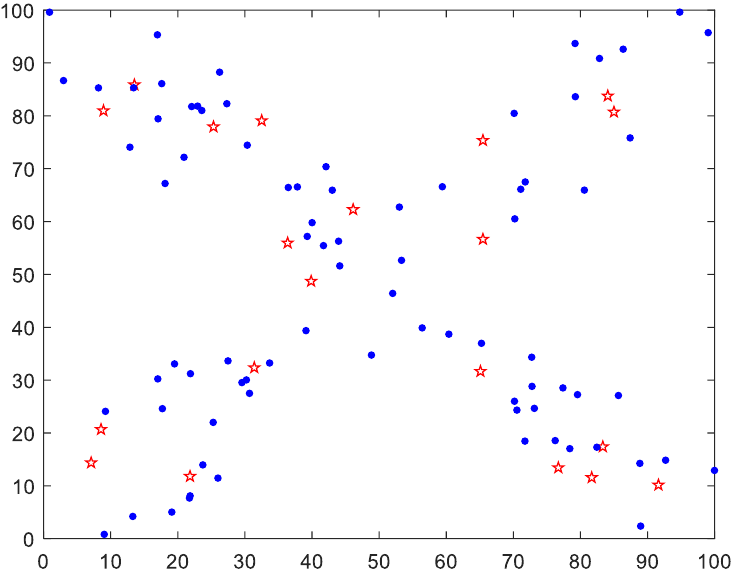}}
	
	\caption{Data sets of different shape test networks. \textcolor{red}{``$ \star $"}: an anchor node, \textcolor{blue}{``$ \bullet $"}: an unknown node. }
	\label{data}
\end{figure}

\begin{table*}[t]
	\renewcommand{\arraystretch}{1.15}
	\caption{The $ ALEs $ of the randomly distributed network in the experiment.}
	\label{randm}
	\centering
	
	\begin{tabular}{ccccccccccccc}
		\hline
		\hline
		\multicolumn{1}{c}{$ {N}_{a} $} & \multicolumn{4}{c}{5} & \multicolumn{4}{c}{10} & \multicolumn{4}{c}{15} \\  
		\cmidrule(lr){1-1}
		\cmidrule(lr){2-5}
		\cmidrule(lr){6-9}
		\cmidrule(lr){10-13}
		
		\multicolumn{1}{c}{ $ R $ (m)} & \multicolumn{1}{c}{25} &  \multicolumn{1}{c}{30} & \multicolumn{1}{c}{35} &  \multicolumn{1}{c}{40} & \multicolumn{1}{c}{25} &  \multicolumn{1}{c}{30} & \multicolumn{1}{c}{35} &  \multicolumn{1}{c}{40} & \multicolumn{1}{c}{25} &  \multicolumn{1}{c}{30} & \multicolumn{1}{c}{35} &  \multicolumn{1}{c}{40} \\
		\hline
		
		DV-Hop\cite{niculescu2003dv09}	&    80.12	&   195.27	&  	142.59	& 122.09& 	38.41	& 	32.75	& 	33.08	& 	30.55	& 	28.65	& 	29.06	& 	32.86	& 	27.09     \\
		
		CC-DV-Hop\cite{gui2020connectivity014}	&    \textbf{40.61} & 122.80  & 83.35  & 85.22 & 34.31 &  30.53 &  29.14 &  27.07 & 23.67 &  27.09 &  28.89 &  23.47\\
		
		OCS-DV-Hop\cite{cui2017novel023} &   50.02 &	71.04& 	76.08& 	76.52& 	31.65& 	26.47& 	30.96& 	26.82& 	24.36& 	21.70& 	27.04& 	23.16    \\
		
		CCS-DV-Hop\cite{8913604@023_1}	&    50.17 &	\textbf{69.12} &	73.54 &	76.49 &	31.65 &	25.51 &	29.34 &	25.60 &	24.35 &	21.28 &	26.40 &	22.82 \\
		
		IAGA-DV-Hop\cite{OUYANG@023_2}	&   102.37	&  92.81	& 82.74	& 75.33 &	30.34 &	27.35  &	29.90  & 26.01 &	24.72 &	21.41 &	25.83 &	21.06  \\
		
		NSGAII-DV-Hop\cite{cai2019multi024} &	91.05 & 	86.90 & 	71.29 & 	63.36 & 	30.14 & 	26.85 & 	29.27 & 	25.60 & 	23.68 & 	22.10 & 	25.32 & 	20.99  \\
		
		DEM-DV-Hop\cite{TVTWang77} &	85.00 & 	82.20 & 	70.03 & 	59.64 & 	27.54 & 	23.72 & 	26.56 & 	24.12 & 	20.77 & 	19.92 & 	22.30 & 	19.30  \\

		\textbf{Ours (DEMN+Hop loss)} & 	80.82 & 	76.98 & 	59.77 & 	\textbf{43.95} & 	\textbf{19.59} & 	\textbf{16.60} & 	\textbf{17.77} & 	\textbf{14.53} & 	\textbf{16.49} & 	\textbf{15.16} & 	\textbf{15.00} & 	\textbf{12.31} 
		\\
		
		\hline

		\textbf{Ours ($ MIter=300 $)} & 87.67 &	82.23 & 	62.67 & 	46.60 & 	22.24 & 	18.42 & 	18.81 & 	15.46 & 	17.76 & 	16.43 & 	15.87 & 	13.03 
		\\

		\textbf{Ours ($ MIter=400 $)} & 	83.03 & 	78.67 & 	60.89 & 	44.76 & 	20.36 & 	17.18 & 	18.14 & 	14.91 & 	16.84 & 	15.49 & 	15.22 & 	12.55 
		\\

		\hline
		
		\textbf{DEMN} & 84.18 & 	80.50 & 	68.10 & 	58.24 & 	24.57 & 	20.82 & 	24.18 & 	22.03 & 	18.15 & 	17.61 & 	19.61 & 	16.49 		\\

		\textbf{Hop loss} & 86.05 &  	80.59 & 	\textbf{58.65} & 	45.49 & 	23.78 & 	18.79 & 	20.35 & 	17.00 & 	18.52 & 	17.09 & 	19.53 & 	16.05 		\\

		\hline
		\hline

		\multicolumn{1}{c}{$ {N}_{a} $} & \multicolumn{4}{c}{20} &  \multicolumn{4}{c}{25} & \multicolumn{4}{c}{30} \\ 
		\cmidrule(lr){1-1}
		\cmidrule(lr){2-5}
		\cmidrule(lr){6-9}
		\cmidrule(lr){10-13}
		
		\multicolumn{1}{c}{ $ R $ (m)} & \multicolumn{1}{c}{25} &  \multicolumn{1}{c}{30} & \multicolumn{1}{c}{35} &  \multicolumn{1}{c}{40} & \multicolumn{1}{c}{25} &  \multicolumn{1}{c}{30} & \multicolumn{1}{c}{35} &  \multicolumn{1}{c}{40} & \multicolumn{1}{c}{25} &  \multicolumn{1}{c}{30} & \multicolumn{1}{c}{35} &  \multicolumn{1}{c}{40}\\
		
		\hline
		
		DV-Hop\cite{niculescu2003dv09}	&    31.98	& 	29.12	& 	28.19	& 	26.16	& 	27.77	& 	27.36	& 	26.38	& 	26.10	& 	33.78	& 	29.63	& 	30.87	& 	26.42     \\
		
		CC-DV-Hop\cite{gui2020connectivity014}	&    26.71  & 24.37  & 23.73 &  22.48 & 22.41  & 23.61 &  21.45 &  22.98 &  27.65 & 26.78 &  25.33 &  20.83\\
		
		OCS-DV-Hop\cite{cui2017novel023} &   24.40	& 	21.91	& 	20.94	& 	22.62	& 	20.42	& 	18.75	& 	19.85	& 	19.02	& 	21.89	& 	18.65	& 	22.27	& 	19.26   \\
		
		CCS-DV-Hop\cite{8913604@023_1}	&    24.25	& 	22.10	& 	20.99	& 	22.13	& 	21.23	& 	19.10 	&	20.16 	&	19.13	& 	22.29	& 	19.09 	&	21.85	& 	19.36  \\
		
		IAGA-DV-Hop\cite{OUYANG@023_2}	&   23.29 &	20.22 &	21.35 &	20.39 &	20.38 &	18.60 &	19.47 &	18.77 &	21.34 &	18.27 &	19.64 &	18.50   \\
		
		NSGAII-DV-Hop\cite{cai2019multi024} &	22.70 & 	21.39 & 	21.06 & 	20.10 & 	19.92 & 	18.94 & 	19.40 & 	18.34 & 	20.48 & 	18.66 & 	19.63 & 	18.30  \\
		
		DEM-DV-Hop\cite{TVTWang77} &	18.59 & 	19.49 & 	19.26 & 	17.78 & 	18.24 & 	17.01 & 	18.32 & 	16.87 & 	18.21 & 	16.78 & 	17.59 & 	16.41   \\

		\textbf{Ours (DEMN+Hop loss)} &	\textbf{16.03} 	& \textbf{13.84} & 	\textbf{13.03} & 	\textbf{11.13} & 	\textbf{14.01} & 	\textbf{12.29} & 	\textbf{11.83} & 	\textbf{10.36} & 	\textbf{14.02} & 	\textbf{12.45} & 	\textbf{11.45} & 	\textbf{9.95} 
		\\

		\hline
		
		\textbf{Ours ($ MIter=300 $)} & 16.95 & 	14.56 & 	13.60 & 	11.92 & 	14.52 & 	12.95 & 	12.37 & 	11.16 & 	14.51 & 	12.86 & 	11.99 & 	10.62 
		\\

		\textbf{Ours ($ MIter=400 $)} & 	16.25 &	14.07 & 	13.20 & 	11.42 & 	14.13 & 	12.44 & 	11.99 & 	10.64 & 	14.15 & 	12.62 & 	11.63 & 	10.21 
		\\

		\hline
		
		\textbf{DEMN}  & 17.66 &	16.81 & 	16.61 & 	15.41 & 	15.73 & 	14.90 & 	15.26 & 	14.57 & 	15.85 & 	14.48 & 	14.87 & 	14.05 		\\

		\textbf{Hop loss}  & 17.59 &	15.26 &	   15.91 &  	15.36 & 	 15.45 &	13.44 &	   14.78 &  	14.01 &	     15.48 &	13.58 &    14.73 &  	13.80		\\

		\hline
		\hline
	\end{tabular}
\end{table*}

\begin{table}[t]
	\renewcommand{\arraystretch}{1.15}
	\caption{Overall performance on randomly distributed networks. }
	\label{rand}
	\centering
	\begin{threeparttable}
		\begin{tabular}{cccc }
			
			\hline
			\hline
			\textbf{ Algorithms} & \textbf{$ ALA $ (\%)}  & \textbf{$ APG $ (\%)}   & \textbf{TC (s)}     \\
			
			\hline
			
			DV-Hop\cite{niculescu2003dv09}     & 70.19    &  15.92 $ \uparrow $  & 0.015\\
			
			OCS-DV-Hop\cite{cui2017novel023}  & 76.89      &  9.22 $ \uparrow $  &  13.76\\
			
			NSGAII-DV-Hop\cite{cai2019multi024}    & 77.86  &  8.25 $ \uparrow $ &  10.95\\
			
			CC-DV-Hop\cite{gui2020connectivity014}  & 74.37   &  11.75 $ \uparrow $ &  197.58\\
			
			CCS-DV-Hop\cite{8913604@023_1}   &  77.36      &  8.75 $ \uparrow $ &  15.09\\
			
			IAGA-DV-Hop\cite{OUYANG@023_2}    & 77.66      &  8.45 $ \uparrow $ &  12.82\\
			
			DEM-DV-Hop\cite{TVTWang77}    & 80.06      &  6.05 $ \uparrow $ &  11.23\\
			
			\textbf{Ours}    & \textbf{86.11}  &  0.00 &  19\\
			
			\hline
			
			\textbf{Ours ($ MIter=300$)}    & 85.20 & 0.91 $ \uparrow $&  11.4\\
			
			\textbf{Ours ($ MIter=400$)}    & 85.83 & 0.28 $ \uparrow $ &  15.2\\
			
			\hline
			
			\textbf{DEMN}  & 82.52 &	3.59 $ \uparrow $& 	11.2	\\

			\textbf{Hop loss}  & 83.47 &	2.64 $ \uparrow $&	  19	\\
			
			\hline
			\hline
		\end{tabular}
		*$ ALA $: average localization accuracy; TC: time consumption.
		
	\end{threeparttable}
\end{table}

\begin{table*}[t]
	\renewcommand{\arraystretch}{1.15}
	\caption{The $ ALEs $ of the C-shaped network in the experiments}
	\label{shapec}
	\centering
	\begin{tabular}{ccccccccccccc}
		\hline
		\hline
		\multicolumn{1}{c}{$ {N}_{a} $} & \multicolumn{4}{c}{5} & \multicolumn{4}{c}{10} & \multicolumn{4}{c}{15} \\  
		\cmidrule(lr){1-1}
		\cmidrule(lr){2-5}
		\cmidrule(lr){6-9}
		\cmidrule(lr){10-13}
		
		$ R $ (m)	&   25	& 	30	& 	35	& 	40	& 	25	& 	30	& 	35	& 	40	& 	25	& 	30	& 	35	& 	40  \\
		\hline
		
		DV-Hop\cite{niculescu2003dv09}	&    88.45	&   120.14	&  	104.47	& 	90.64	& 	66.83	& 	93.20	& 	81.80	& 	66.76	& 	67.79	& 	91.93	& 	76.57   & 	70.52     \\
		
		CC-DV-Hop\cite{gui2020connectivity014}	&    83.98 & 73.33  & 64.06  & 63.87 & 52.25 &  47.20  &  46.30 &  38.77 & 52.76 &  50.12 &  43.18 &  45.59\\
		
		OCS-DV-Hop\cite{cui2017novel023} &   \textbf{47.77} &	\textbf{40.02}   & 	37.57   & 	36.71   & 	42.26   & 	35.14   & 	35.29   & 35.48   & 	45.67   & 	36.35   & 	34.05   & 	33.46    \\
		
		CCS-DV-Hop\cite{8913604@023_1}&   50.09 &	41.14 &	38.41	&   34.29  &	42.97 &	32.88 &	33.34 &	33.02 &	44.90   &	36.29	&	33.24 &	32.96 \\
		
		IAGA-DV-Hop\cite{OUYANG@023_2}	&   116.33	&  103.63	& 86.13	& 78.20 &	41.76 &	32.55  &	32.97  & 31.71 &	42.97 &	34.98 &	32.95 &	33.11 \\
		
		NSGAII-DV-Hop\cite{cai2019multi024} &	69.82	& 	58.85	& 	58.03	& 	54.38	& 	39.33	& 	31.76	& 	28.24	& 	28.91	& 	40.23	& 	31.26	& 	27.23	& 	28.72  \\
		
		DEM-DV-Hop\cite{TVTWang77} &	62.34  &    53.11 & 	56.83 & 	51.81 &    40.21    & 	31.55 & 	27.69 & 	27.37 & 	38.44 & 	30.04 & 	26.35 & 	26.99 \\

		\textbf{Ours} &	57.69 & 	43.35 & 	\textbf{35.78} & 	\textbf{33.37} & 	\textbf{34.55} & 	\textbf{27.32} & 	\textbf{21.63} & 	\textbf{21.21} & 	\textbf{33.99} & 	\textbf{27.64} & 	\textbf{22.86} & 	\textbf{21.95} 
		\\ 
		
		\hline
		\hline
		
		\multicolumn{1}{c}{$ {N}_{a} $} & \multicolumn{4}{c}{20} & \multicolumn{4}{c}{25} & \multicolumn{4}{c}{30} \\ 
		\cmidrule(lr){1-1}
		\cmidrule(lr){2-5}
		\cmidrule(lr){6-9}
		\cmidrule(lr){10-13}
		
		$ R $ (m)	&   25	& 	30	& 	35	& 	40	& 	25	& 	30	& 	35	& 	40	& 	25	& 	30	& 	35	& 	40  \\
		\hline
		
		DV-Hop\cite{niculescu2003dv09}	&    61.31	& 	68.12	& 	59.57	& 	52.60	& 	60.91	& 	59.27	& 	53.28	& 	46.08	& 	63.28	& 	53.51	& 	49.10	& 	42.31     \\
		
		CC-DV-Hop\cite{gui2020connectivity014}	&    43.91  & 38.03  & 41.07 &  37.84 & 44.02  & 39.50 &  40.02 &  34.51 &  44.88 & 40.87 & 42.16 &  35.14\\
		
		OCS-DV-Hop\cite{cui2017novel023} &   42.61	& 	35.09	& 	35.85	& 	32.45	& 	54.28	& 	39.17	& 	40.97	& 36.02	& 	53.55	& 	40.38	& 	42.64	& 	37.98   \\
		
		CCS-DV-Hop\cite{8913604@023_1}	&    43.65	& 	34.97	& 	35.51	& 	32.39	& 	50.81	& 	37.29 	&	39.86 	&	34.16	& 	53.97	& 	38.72	&	42.37	& 	37.79  \\
		
		IAGA-DV-Hop\cite{OUYANG@023_2}	&   35.59 &	29.70 &	29.84 &	28.84 &	39.66 &	31.05 &	31.81 &	29.17 &	38.63 &	30.95 &	31.99 &	28.93   \\
		
		NSGAII-DV-Hop\cite{cai2019multi024} &	36.04  &	30.44 & 27.23 & 	27.17 & 	38.23 & 	30.83 & 28.11 & 	27.21  &	37.17 & 	31.52 & 	27.87 & 	26.32   \\
		
		DEM-DV-Hop\cite{TVTWang77}   &   35.44  &	29.32   &	25.34   &	26.04  &	37.23   &	29.52  & 	25.96   &   25.89   &	35.98   &	29.60  &	25.84   &   25.23  \\

		\textbf{Ours}   &   \textbf{31.00} & 	\textbf{25.54} & 	\textbf{21.31} & 	\textbf{20.60} & 	\textbf{30.96} & 	\textbf{24.98} & 	\textbf{21.09} & 	\textbf{20.04} & 	\textbf{29.38} & 	\textbf{25.02} & 	\textbf{20.87} & 	\textbf{19.21} 
		\\ 
		
		\hline
		\hline
	\end{tabular}
\end{table*}

\begin{table*}[t]
	\renewcommand{\arraystretch}{1.15}
	\caption{The $ ALEs $ of the O-shaped network in the experiments}
	\label{shapeo}
	\centering
	\begin{tabular}{ccccccccccccc}
		\hline
		\hline
		\multicolumn{1}{c}{$ {N}_{a} $} & \multicolumn{4}{c}{5} & \multicolumn{4}{c}{10} & \multicolumn{4}{c}{15} \\  
		\cmidrule(lr){1-1}
		\cmidrule(lr){2-5}
		\cmidrule(lr){6-9}
		\cmidrule(lr){10-13}
		
		$ R $ (m)	&   25	& 	30	& 	35	& 	40	& 	25	& 	30	& 	35	& 	40	& 	25	& 	30	& 	35	& 	40  \\
		\hline
		
		DV-Hop\cite{niculescu2003dv09}	&    84.04	&   152.28	&  	121.71	& 	100.22	& 	50.68	& 	136.31	& 	109.29	& 	92.96	& 	53.49	& 	101.91	& 	80.13	& 	68.32     \\
		
		CC-DV-Hop\cite{gui2020connectivity014}	&    \textbf{41.56} & \textbf{35.11}  & 46.24  & \textbf{36.64} & 42.70 &  27.78 &  24.31 &  34.78 & 38.49 &  28.15 &  21.56 &  26.27\\
		
		OCS-DV-Hop\cite{cui2017novel023} &   43.29 &    46.62   & 	43.39   & 	42.82   & 	37.64   & 	34.66   & 	31.86   & 	28.95   & 	38.10   & 	30.04   & 	26.93   & 	24.01    \\
		
		CCS-DV-Hop\cite{8913604@023_1}	&    41.80 &	45.16 &	43.57 &	42.69 &	37.74 &	33.09 &	30.36 &	27.84 &	38.07 &	30.02 &	27.32 &	23.97 \\
		
		IAGA-DV-Hop\cite{OUYANG@023_2}	&   62.26	&  61.53	& 56.40	& 51.91 &	34.49 &	30.54  &	26.55  & 29.53 &	30.79 &	24.50 &	21.08 &	21.77  \\
		
		NSGAII-DV-Hop\cite{cai2019multi024} &	55.11	& 	59.35	& 	46.04	& 	47.87	& 	35.77	& 	27.84	& 	24.62	& 	27.98	& 	32.53	& 	22.33	& 	20.59	& 	20.24  \\
		
		DEM-DV-Hop\cite{TVTWang77} &	50.72  &	52.84 & 	42.33 & 	44.42 &    33.02 & 	24.85 & 	21.96 & 	22.57 & 	32.57 & 	21.42 & 	18.70 & 	17.49 \\ 
		
		\textbf{Ours} &	 75.45 & 	56.14 & 	\textbf{40.80} & 	37.33 & 	\textbf{32.43} & 	\textbf{21.29} & 	\textbf{18.34} & 	\textbf{17.14} & 	\textbf{30.67} & 	\textbf{19.81} & 	\textbf{16.93} & 	\textbf{14.28} 
		\\ 
		
		\hline
		\hline
		
		\multicolumn{1}{c}{$ {N}_{a} $} & \multicolumn{4}{c}{20} & \multicolumn{4}{c}{25} & \multicolumn{4}{c}{30} \\ 
		\cmidrule(lr){1-1}
		\cmidrule(lr){2-5}
		\cmidrule(lr){6-9}
		\cmidrule(lr){10-13}
		
		$ R $ (m)	&   25	& 	30	& 	35	& 	40	& 	25	& 	30	& 	35	& 	40	& 	25	& 	30	& 	35	& 	40  \\
		\hline
		
		DV-Hop\cite{niculescu2003dv09}	&    47.14	& 	92.79	& 	73.85	& 	64.74	& 	48.10	& 	83.41	& 	67.19	& 	57.83	& 	48.06	& 	47.54	& 	44.59	& 	37.29     \\
		
		CC-DV-Hop\cite{gui2020connectivity014}	&    37.39  & 27.11  & 24.26 &  26.05 & 38.29  & 26.88 &  21.37 &  23.35 &  38.77 & 24.99 &  27.43 &  24.28\\
		
		OCS-DV-Hop\cite{cui2017novel023} &   40.07	& 	34.53	& 	28.52	& 	24.77	& 	46.46	& 	35.16	& 	30.84	& 	24.87	& 	48.43	& 	38.93	& 	32.48	& 	28.96   \\
		
		CCS-DV-Hop\cite{8913604@023_1}	&    40.23	& 	33.09	& 	27.53	& 	23.95	& 	46.32	& 	34.70 	&	30.14 	&	24.97	& 	50.08	& 	36.90 	&	32.00	& 	26.80  \\
		
		IAGA-DV-Hop\cite{OUYANG@023_2}	&   29.90 &	23.26 &	20.35 &	19.15 &	32.32 &	22.45 &	20.43 &	18.94 &	31.53 &	22.25 &	21.01 &	19.00   \\
		
		NSGAII-DV-Hop\cite{cai2019multi024} &	30.11  &	21.33 & 	18.92 & 	17.37 & 	31.92 & 	21.22 & 19.10 & 	17.83  &	30.26 & 	20.57 & 	19.76 & 	17.94   \\
		
		DEM-DV-Hop\cite{TVTWang77}   &	29.25  &	20.87  &	17.24  &	14.92  &	30.73  &	21.03  & 	16.83  &	14.51  &	28.90  &	20.09  &	17.06  &	15.16  \\

		\textbf{Ours}  &  \textbf{28.16} &	\textbf{18.71} & 	\textbf{15.60} & 	\textbf{12.93} & 	\textbf{27.63} & 	\textbf{17.91} & 	\textbf{14.98} & 	\textbf{12.34} & 	\textbf{25.24} & 	\textbf{17.18} & 	\textbf{14.95} & 	\textbf{12.41}
		\\

		\hline
		\hline
	\end{tabular}
\end{table*}

\begin{table*}[t]
	\renewcommand{\arraystretch}{1.15}
	\caption{The $ ALEs $ of the X-shaped network in the experiments}
	\label{shapex}
	\centering
	\begin{tabular}{ccccccccccccc}
		\hline
		\hline
		\multicolumn{1}{c}{$ {N}_{a} $} & \multicolumn{4}{c}{5} & \multicolumn{4}{c}{10} & \multicolumn{4}{c}{15} \\  
		\cmidrule(lr){1-1}
		\cmidrule(lr){2-5}
		\cmidrule(lr){6-9}
		\cmidrule(lr){10-13}
		
		$ R $ (m)	&   25	& 	30	& 	35	& 	40	& 	25	& 	30	& 	35	& 	40	& 	25	& 	30	& 	35	& 	40  \\
		\hline
		
		DV-Hop\cite{niculescu2003dv09}	&    58.46	&   108.88	&  	95.24	& 	91.84	& 	60.38	& 	96.73	& 	86.75	& 	83.64	& 	51.12	& 	81.10	& 	70.13	& 	72.57     \\
		
		CC-DV-Hop\cite{gui2020connectivity014}	&    52.30 & 49.44  & 49.42  & 45.36 & 55.27 &  46.77 &  45.76 &  40.71 & 43.97 &  38.37 &  35.97 &  36.14\\
		
		OCS-DV-Hop\cite{cui2017novel023} &   42.85 &	49.13   & 	47.49   & 	38.15   & 	39.75   & 	40.24   & 	35.80   & 	32.52   & 	42.18   & 	41.91   & 	37.48   & 	35.86    \\
		
		CCS-DV-Hop\cite{8913604@023_1}	&    42.69  &	48.68   &	45.04   &	34.45   &	40.18   &	40.08   &	34.36   &	30.64   &	38.37   &	38.90   &	33.59   &	32.61 \\
		
		IAGA-DV-Hop\cite{OUYANG@023_2}	&   47.46	&  49.47	& 43.92	& 40.17 &	33.99 &	35.08  &	31.10   & 30.07 &	30.83   &	29.51   &	26.14 &	27.25  \\
		
		NSGAII-DV-Hop\cite{cai2019multi024} &	42.37	& 	44.45	& 	38.87	& 	34.70	& 	33.27	& 	31.89	& 	27.84	& 	27.18	& 	29.96	& 	28.18	& 	23.68	& 	24.20  \\
		
		DEM-DV-Hop\cite{TVTWang77} &	40.43  &	42.36 & 	37.50 & 	33.24 &    32.95 & 	30.90 & 	26.03 & 	25.14 & 	29.39 & 	27.26 & 	21.81 & 	21.27 \\

		\textbf{Ours} &	\textbf{32.46} &	 \textbf{33.69} &	\textbf{30.59} & 	\textbf{25.42} & 	\textbf{28.13} & 	\textbf{23.95} & 	\textbf{19.30} & 	\textbf{18.38} & 	\textbf{26.36} & 	\textbf{23.07} & 	\textbf{17.76} & 	\textbf{17.24} 
		\\

		\hline
		\hline
		
		\multicolumn{1}{c}{$ {N}_{a} $} & \multicolumn{4}{c}{20} & \multicolumn{4}{c}{25} & \multicolumn{4}{c}{30} \\ 
		\cmidrule(lr){1-1}
		\cmidrule(lr){2-5}
		\cmidrule(lr){6-9}
		\cmidrule(lr){10-13}
		
		$ R $ (m)	&   25	& 	30	& 	35	& 	40	& 	25	& 	30	& 	35	& 	40	& 	25	& 	30	& 	35	& 	40  \\
		\hline
		
		DV-Hop\cite{niculescu2003dv09}	&    49.07	& 	78.38	& 	68.07	& 	67.90	& 	52.52	& 	69.58	& 	61.29	& 	56.98	& 	54.51	& 	56.82	& 	49.58	& 	47.62     \\
		
		CC-DV-Hop\cite{gui2020connectivity014}	&    42.36  & 40.13  & 36.76 &  38.06 & 45.50  & 44.27 &  40.72 &  39.52 &  45.83 & 44.09 &  39.54 &  39.61\\
		
		OCS-DV-Hop\cite{cui2017novel023} &   43.51	& 	45.41	& 	37.52	& 	36.84	& 	45.50	& 	45.87	& 	39.09	& 	42.20	& 	45.51	& 	49.90	& 	42.38	& 	45.61   \\
		
		CCS-DV-Hop\cite{8913604@023_1}	&    42.11	& 	43.44	& 	35.62	& 	36.04	& 	46.39	& 	46.22 	&	38.32 	&	40.86	& 	44.27	& 	48.19 	&	41.28	& 	44.55  \\
		
		IAGA-DV-Hop\cite{OUYANG@023_2}	&   30.42 &	29.60 &	27.35 &	27.01 &	29.89 &	28.34 &	26.35 &	26.90 &	26.34 &	26.89 &	25.70 &	25.96   \\
		
		NSGAII-DV-Hop\cite{cai2019multi024} &	27.94  &	26.55 & 	23.04 & 	23.18 & 	28.38 & 	26.49 & 22.24 & 	23.18  &	25.18 & 	24.32 & 	20.91 & 	20.91  \\
		
		DEM-DV-Hop\cite{TVTWang77}   &	27.70  &	25.03  &	21.65  &	20.96  &	28.10  &	25.47  & 	20.23  &	20.52  &	25.05  &	23.51  &	18.81  &	18.02  \\
		
		\textbf{Ours}   &	\textbf{23.85} & 	\textbf{20.91} & 	\textbf{17.38} & 	\textbf{17.06} & 	\textbf{24.65} & 	\textbf{21.15} & 	\textbf{16.51} & 	\textbf{16.91} & 	\textbf{22.82} & 	\textbf{19.68} & 	\textbf{14.47} & 	\textbf{15.07} 
		\\ 
		
		\hline
		\hline
	\end{tabular}
\end{table*}

\begin{table*}[t]
	\renewcommand{\arraystretch}{1.15}
	\caption{A comprehensive result comparison. }
	\label{Table}
	\centering
	\begin{threeparttable}
		\begin{tabular}{cccccccc}
			
			\hline
			\hline
			\multicolumn{1}{c}{Node distribution type} & \multicolumn{2}{c}{C-shaped networks} & \multicolumn{2}{c}{O-shaped networks} & \multicolumn{2}{c}{X-shaped networks} \\ 
			
			\cmidrule(lr){1-1}
			\cmidrule(lr){2-3}
			\cmidrule(lr){4-5}
			\cmidrule(lr){6-7}
			
			\textbf{ Algorithms} & \textbf{$ ALA $ (\%)}  & \textbf{$ APG $ (\%)}   & \textbf{$ ALA $ (\%)}  & \textbf{$ APG $ (\%)}   & \textbf{$ ALA $ (\%)}  & \textbf{$ APG $ (\%)}   &     \\
			
			\hline

			OCS-DV-Hop\cite{cui2017novel023}  & 57.10      &  17.84  $ \uparrow $ &  70.79  &  9.76 $ \uparrow $  &   58.03  &  21.74 $ \uparrow $ \\
			
			NSGAII-DV-Hop\cite{cai2019multi024}    & 60.57  &  14.37  $ \uparrow $&  66.69   &  13.86 $ \uparrow $  &   58.75  &   21.02 $ \uparrow $\\
			
			CC-DV-Hop\cite{gui2020connectivity014}  & 61.45   &  13.49 $ \uparrow $ &  67.24   &  13.31 $ \uparrow $  &   60.20  &   19.57 $ \uparrow $\\
			
			CCS-DV-Hop\cite{8913604@023_1}   &  66.54      &  8.40  $ \uparrow $&  75.01   &  5.54 $ \uparrow $  &   71.26  &  8.51 $ \uparrow $ \\
			
			IAGA-DV-Hop\cite{OUYANG@023_2}    & 68.81      &  6.13  $ \uparrow $&  76.09   &   4.46 $ \uparrow $ &  74.07   &  5.70 $ \uparrow $ \\
			
			DEM-DV-Hop\cite{TVTWang77}    & 70.00    &  4.94 $ \uparrow $ &  78.04   &  2.51 $ \uparrow $  &  75.51   &  4.26 $ \uparrow $ \\
			
			\hline
			
			\textbf{Ours}    & \textbf{74.94}  &  0.00 &  \textbf{80.55}  &  0.00  &  \textbf{79.77}  &  0.00  \\

			\hline
			\hline
		\end{tabular}
		
	\end{threeparttable}
\end{table*}

\begin{figure}[t]
	\center
	\subfigure[]{\includegraphics[scale=0.34]{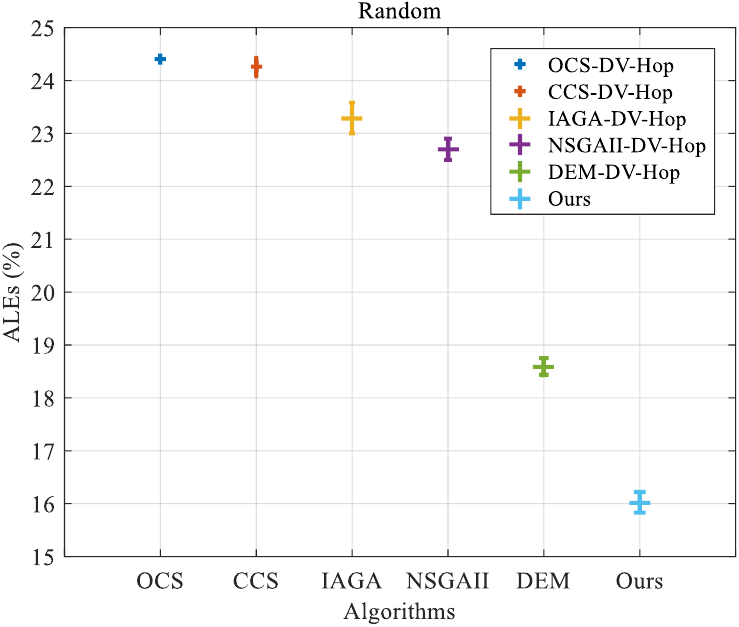}}
	\subfigure[]{\includegraphics[scale=0.34]{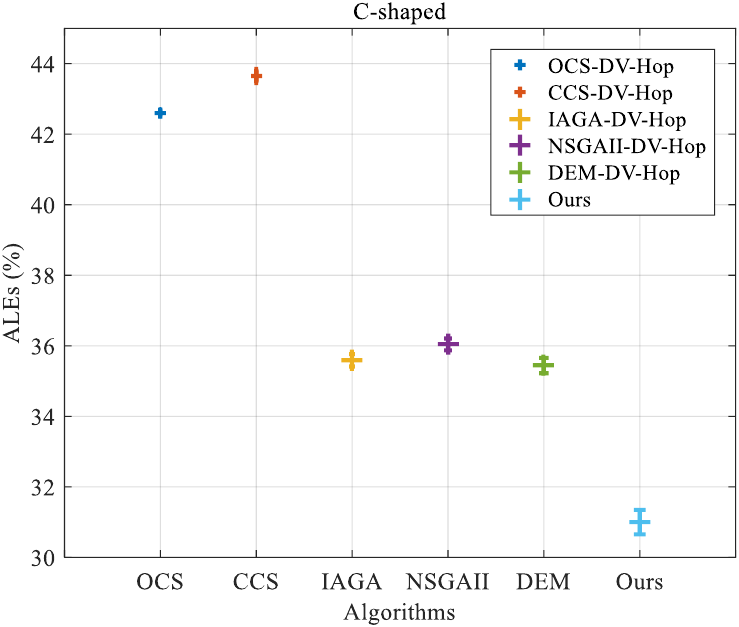}}
	\subfigure[]{\includegraphics[scale=0.34]{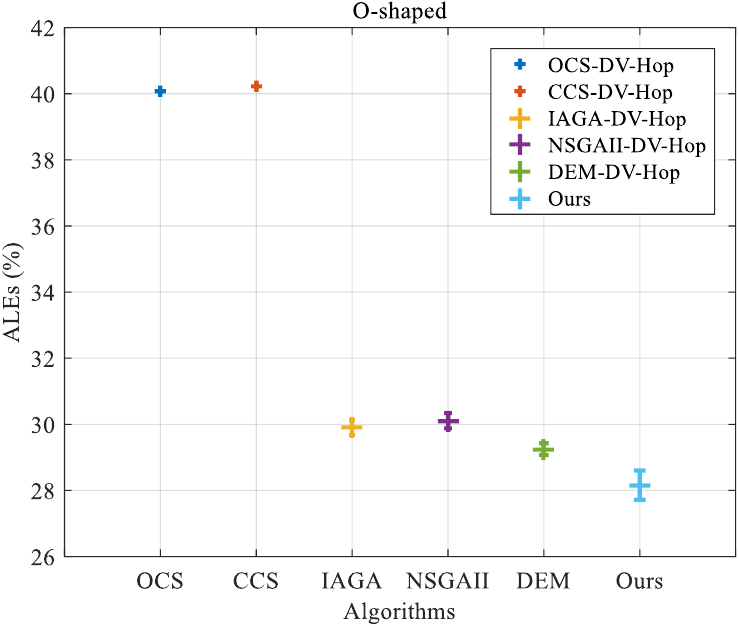}}
	\subfigure[]{\includegraphics[scale=0.34]{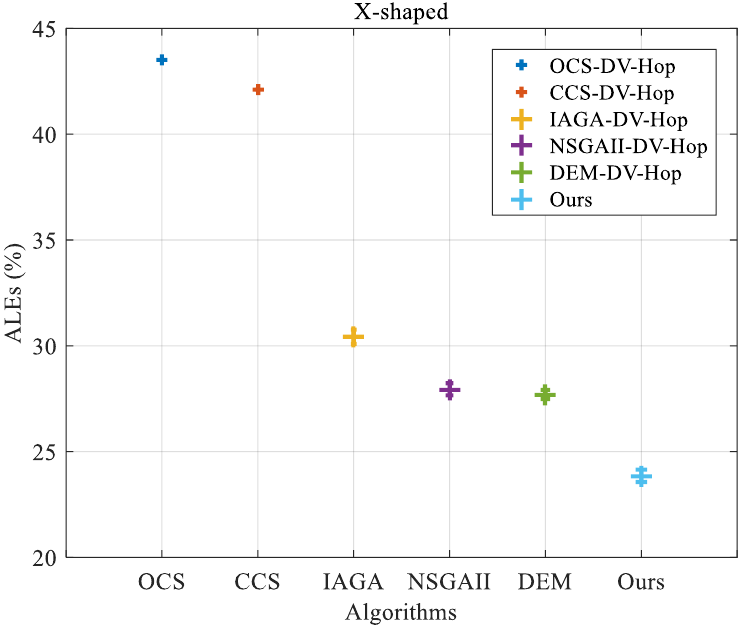}}
	
	\caption{The $ 95\% $ confidence interval for $ ALEs $ samples. (a): randomly distributed network; (b): C-shaped network; (c): O-shaped network; (d): X-shaped network. }
	\label{std}
\end{figure}

\iffalse
\begin{figure}[htbp]
	
	\center
	\subfigure[]{\includegraphics[scale=0.45]{7_1.pdf}}
	\subfigure[]{\includegraphics[scale=0.45]{7_2.pdf}}
	
	\caption{The convergence speed of the proposed model. $ {N}_{a} $: the number of anchor nodes.}
	\label{converg}
\end{figure}
\fi

\textbf{Evaluation Metrics.} The normalized average localization errors ($ ALEs $) represent the most commonly used metric for evaluating localization, while the average performance gain ($ APG $) indicates the improvement of the proposed model in comparison to other algorithms. In this paper, we employ both to assess our model's reliability and robustness. The $ ALEs $ are calculated by
\begin{equation}
ALEs = \frac{100\%}{{N_{u} \cdot R}}\sum\limits_{k = 1}^{N_{u}} {\sqrt {{{(x_k^{gt} - {x_k})}^2} + {{(y_k^{gt} - {y_k})}^2}} }   , 
\label{eq18}
\end{equation}
where $ (x_k,y_k) $ and $ (x_k^{gt},y_k^{gt}) $ represent the expected coordinates and ground truth coordinates of an $ u_k $ respectively. And the $ APG $ is calculated by
\begin{equation}
APG= {\rm mean}\sum{({ ALEs}_{other}-{ALEs}_{our})}  , 
\label{eq15}
\end{equation}
where $ {ALEs}_{other} $ represent the $ ALEs $ of other algorithms used for comparison in this paper. In addition, we provide confidence interval (where, the confidence probability is $ 95\% $) analysis for $ ALEs $. Assuming $ ALEs $ follow a Gaussian distribution, the confidence interval is calculated by
\begin{equation}
[\bar{X}-\frac{S}{\sqrt n}t_{{\frac{\alpha}{2}}}(n-1), \bar{X}+\frac{S}{\sqrt n}t_{{\frac{\alpha}{2}}}(n-1)], 
\label{eqint}
\end{equation}
where $ \bar{X} $ denotes the sample mean, $ S^2 $ denotes the sample variance, $ n $ denoites the number of samples, and $ \alpha $ denotes the significance level. In this work, $ \bar{X}={\frac{1}{n}}\sum_{i}^{n}{ALE}_i $, $ S^2=\frac{\sum_{i}^{n}{({ALE}_i-\bar{X})}^2}{n-1} $, $ n=50 $, and $\alpha =0.05 $.

\subsection{Experimental Results}

Table \ref{randm} compares the performance with the existing excellent work under a randomly distributed network. When $ N_a>5 $, the model proposed in this paper obtains the minimum $ ALEs $. When $ N_a=5 $, our algorithm also shows strong competitiveness. When $ R=25 $ or $ 30 $, the $ ALEs $ of the proposed model are greater than CC-DV-Hop\cite{gui2020connectivity014}, OCS-DV-Hop\cite{cui2017novel023} and CCS-DV-Hop\cite{8913604@023_1}, and the minimum $ ALEs $ are maintained in other cases. When $ N_a=30 $ and $ R=40 $, the $ ALEs $ of the proposed model are less than $ 10\% $. In addition, Table \ref{randm} also provides $ ALEs $ of the proposed model when $ MaxIter=300 $ or $ 400 $, and the results show that the proposed model still achieves the best localization accuracy. Additionally, we conduct an ablation study to demonstrate the isolated impacts of each design component within our proposed approach. We illustrate how proposed DEMN in section III and the hop loss in section IV enhance the baselines (i.e., DEM-DV-Hop\cite{TVTWang77}). We utilize the identical solving algorithm as the baselines\cite{TVTWang77}, which can more convincingly demonstrate the advantages of the proposed model. The results suggest that replacing the distance estimation model with DEMN in the baselines results in a reduction of $ 0.82\%-3.06\% $ in $ ALEs $ compared to the baselines, while increasing APG by $ 2.46\% $. Correspondingly, the localization performance achieved solely through the hop loss surpasses that of the baselines and only falls short when $ {N}_{a}=5 $ and $ R=25 $ are considered. Its $ APG $ and maximum performance gain are $ 3.41\% $ and $ 14.15\% $, respectively. When combining DEMN with the hop loss, we observed a decrease of $ 2.56\%-15.69\% $ in $ ALEs $ compared to baselines, with an $ APG $ of $ 6.05\% $. It is worth noting that both proposed models contribute to improving localization performance, thus validating their effectiveness.

Table \ref{rand} shows the overall performance comparison on the randomly distributed network. The results show that when $ MIter=500$, the average localization accuracy ($ ALA=1-mean(ALEs)$) of the proposed model is up to $ 86.11\% $, which is $ 15.92\% $ higher than the classical DV-Hop\cite{niculescu2003dv09}  algorithm. Compared with the current best performance DEM-DV-Hop\cite{TVTWang77}, the $ APG $ of the proposed model is up to $ 6.05\% $. In addition, Table \ref{rand} also provides the time consumption of each algorithm. When $ MIter=500$, the time consumption of the proposed model is $ 3.91-8.05s $ more than those of the excellent current algorithm \cite{cui2017novel023, cai2019multi024, 8913604@023_1, TVTWang77, OUYANG@023_2}. When $ MIter=300$, the time consumption of the proposed model is equivalent to or even smaller than those of these comparison algorithms. Our model still maintains significant performance advantages in this case, with an $ APG $ of $ 5.14\% - 10.84\% $.

Table \ref{shapec} compares the performance with the existing excellent work under the C-shaped network. The localization accuracy of the proposed model is better than that of the comparison algorithm in most cases. Only when $ N_a=5 $ and $ R=25 $ or $ 30 $, the localization accuracy is inferior to those of OCS-DV-Hop\cite{cui2017novel023} and CCS-DV-Hop\cite{8913604@023_1}. Compared with the NSGAII-DV-Hop\cite{cai2019multi024}, the $ ALEs $ are reduced by $ 3.62\%-22.26\% $, and compared with the DEM-DV-Hop\cite{TVTWang77}, the $ ALEs $ are reduced by $ 2.4\%-21.05\% $.

Table \ref{shapeo} compares the performance with the existing excellent work under the O-shaped network. When $ N_a>5 $, the $ ALEs $ of the proposed model are significantly smaller than those of all comparison algorithms. When $ N_a=5 $ and $ R=25 $ or $ 30 $, the localization performance of the proposed model is inferior to those of CC-DV-Hop\cite{gui2020connectivity014},  OCS-DV-Hop\cite{cui2017novel023}, CCS-DV-Hop\cite{8913604@023_1} and DEM-DV-Hop\cite{TVTWang77}. When $ N_a=5 $ and $ R=40 $, the $ ALEs $ of the proposed model are more significant than that of CC-DV-Hop\cite{gui2020connectivity014}, but less than those of other comparison algorithms. When $ N_a=5 $ and $ R=35 $, the proposed model has the minimum $ ALEs $ compared with these state-of-the-art algorithms.

Table \ref{shapex} compares the performance with the existing excellent work under the X-shaped network. The results show that the $ ALEs $ of the proposed model are significantly lower than other comparison algorithms in all cases. Compared with  NSGAII-DV-Hop\cite{cai2019multi024} and DEM-DV-Hop\cite{TVTWang77}, the localization accuracy of the proposed model is improved by $ 2.35\%-10.77\% $ and $ 2.23\%-8.67\% $ respectively.

Table \ref{Table} shows the $ ALA $ of the proposed model on C-, O- and X-shaped networks and its $ APG $ compared with other algorithms. In this comparison, we have opted not to include the DV-Hop\cite{niculescu2003dv09} algorithm and the test data when $ N_a=5 $ due to some algorithms having $ ALEs $ exceeding $ 100\% $, rendering the results meaningless. The results indicate that the proposed model achieves $ ALAs $ of $ 74.94\% $, $ 80.55\% $, and $ 79.77\% $ on C-, O- and X-shaped datasets, respectively. Compared to the comparison algorithm, our model obtains the $ APGs $ ranging from $ 4.94\%-17.84\% $, $ 2.51\%-13.86\% $, and $ 4.26\%-21.74\% $, respectively.

Fig. \ref{ALA} visually shows the $ ALAs $ of different algorithms on four test datasets. As can be seen in Fig.\ref{ALA}, under the randomly distributed network, the $ ALA $ of NSGAII-DV-Hop\cite{cai2019multi024} outperforms  that of OCS-DV-Hop\cite{cui2017novel023} and CC-DV-Hop\cite{gui2020connectivity014}. However, under the C-, O- and X-shaped networks, the $ ALA $ of NSGAII-DV-Hop\cite{cai2019multi024} is inferior to that of CC-DV-Hop\cite{gui2020connectivity014}. In contrast, our model outperforms other algorithms with the highest $ ALA $ in four data sets, demonstrating its robustness and generalization ability.

Fig. \ref{std} provides the $ 95\% $ confidence intervals (CI) for the samples of  $ ALEs $ obtained under the conditions of $ N_a=20 $ and $ R=25 $.  Factors influencing the width of the CI consist of the confidence level, the dispersion in the sample, and the sample size. For each algorithm investigated in this paper, the sample size and confidence level remain consistent throughout the analysis. Therefore, the greater the dispersion, the wider the confidence interval. The findings demonstrate that across the four distinct test networks, the overall distribution of CI is smaller than comparison algorithms. The width of the CI is slightly wider than the comparison algorithm, i.e. the dispersion of $ ALEs $ samples is slightly higher than comparison algorithms.

\section{Conclusions}

In this paper, to address the following two issues: 1) how to sufficiently utilize the connection information between multiple nodes and 2) how to select a suitable solution from multiple solutions obtained by Euclidean distance loss, we propose a DV-Hop localization based on the distance estimation using multinode (DEMN) and the hop loss in wireless sensor networks. In DEMN, when multiple anchor nodes can detect an unknown node, the distance expectation between the unknown node and an anchor node is calculated using the cross domain information and is considered as the expected distance between them. Compared to the distance estimation based on a single anchor node, DEMN narrows the search space, resulting in more accurate distance estimation. When minimizing the traditional Euclidean distance loss, multiple solutions may exist. To select a suitable solution, the hop loss is proposed, which minimizes the difference between the real and its predicted hops. Finally, the Euclidean distance loss calculated by DEMN and the hop loss are embedded into the multi-objective optimization algorithm. The experimental results show that the proposed method gains 86.11\% location accuracy in the randomly distributed network, which is 6.05\% better than the DEM-DV-Hop, while DEMN and the hop loss can contribute 2.46\% and 3.41\%, respectively. Moreover, experimental results on C-, O-, and X-shaped networks show that the proposed method significantly improves the localization performance.

In future work, we will study DEMN when an unknown node is detected by more than two anchor nodes, while saving computational costs. In addition, we will also consider the calculation method of the expected distance when the nodes in the cross domain follow different distributions.

\ifCLASSOPTIONcaptionsoff
  \newpage
\fi

\bibliographystyle{IEEEtran}
\bibliography{refs}


\end{document}

\appendices
\section{The calculation process of $ {E\_dis}_{i,k} $ when $ m=2 $ and $ d_{i,j}<UB_{i,m} $}

\begin{figure}[htbp]
	
	\centering
	{\includegraphics[scale=0.8]{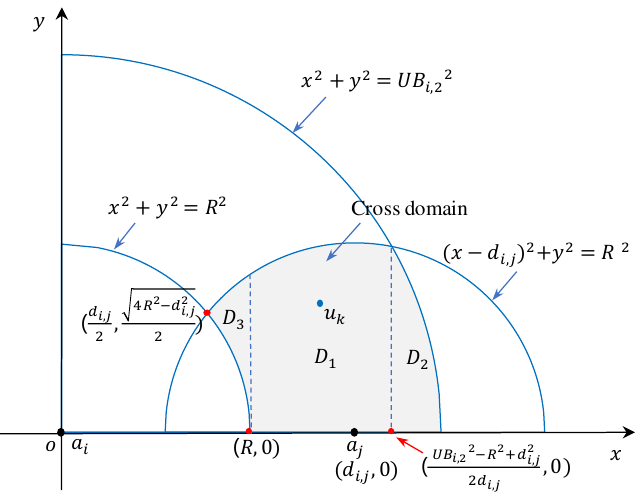}}
	%{\includegraphics[scale=0.8]{2_2.pdf}}
	\caption{Multinode joint probability distribution model when $ m=2 $ and $ R<d<UB_m $. $ {a}_{i} $: an anchor node, $ {u} $: an unknown node, $ D_i $: a part of the cross domain, $ m $: the hop count, $ R $: communication radius,  $ d $: the distance between node $ {a}_{1} $ to node $ {a}_{2} $. }
	\label{joint2}
\end{figure}

When $ m=2 $ and $ d_{i,j}<UB_m $, the visual analysis of the MDE is shown in Fig. \ref{joint2}. The $ {E\_dis}_{i,k} $ between the node $ u_k $ and $ a_i  $ is calculated by
\begin{equation}
\begin{aligned}
{E\_dis}_{i,k}&=\frac{\int_{d_{i,j}-R}^{\frac{{\ {UB}_{i,2}}^2-R^2+d_{i,j}^2}{2d_{i,j}}}{\int_{0}^{\sqrt{R^2-{(x-d_{i,j})}^2}}\sqrt{x^2+y^2}dydx}}{{D_1}}
\\
&+\frac{\int_{\frac{{\ {UB}_{i,2}}^2-R^2+d_{i,j}^2}{2d_{i,j}}}^{{UB}_{i,2}}{\int_{0}^{\sqrt{{{UB}_{i,2}}^2-x^2}}\sqrt{x^2+y^2}dydx}}{{D_2}}
\\
&+\frac{\int_{\frac{d_{i,j}}{2}}^{R}{\int_{\sqrt{R^2-x^2}}^{\sqrt{R^2-{(x-d_{i,j})}^2}}\sqrt{x^2+y^2}dydx}}{{D_3}}
\end{aligned},
\label{eq3}
\end{equation}
where $ d_{i,j} $ denotes the distance between the node $ a_j $ and $ a_i $. $ {D_1} $, $ {D_2} $ and ${D_3} $ are calculated by
\begin{equation}
\begin{aligned}
{D_1}=\int_{d_{i,j}-R}^{\frac{{\ {UB}_{i,2}}^2-R^2+d_{i,j}^2}{2d_{i,j}}}{\sqrt{R^2-{(x-d_{i,j})}^2}dx} 
\end{aligned}, 
\label{eq22}
\end{equation}
\begin{equation}
\begin{aligned}
{D_2}=\int_{\frac{{\ {UB}_{i,2}}^2-R^2+d_{i,j}^2}{2d_{i,j}}}^{{UB}_{i,2}}{\sqrt{{{UB}_{i,2}}^2-x^2}dx}
\end{aligned}, 
\label{eq22}
\end{equation}
\begin{equation}
\begin{aligned}
{D_3}=\int_{\frac{d_{i,j}}{2}}^{R}{(\sqrt{R^2-{(x-d_{i,j})}^2}-\sqrt{R^2-x^2})dx}
\end{aligned}.
\label{eq22}
\end{equation}

\bibliographystyle{IEEEtran}
